\documentclass[Physsubmission, Phys]{SciPost}

\hypersetup{
    colorlinks,
    linkcolor={red!50!black},
    citecolor={blue!50!black},
    urlcolor={blue!80!black}
}

\usepackage[bitstream-charter]{mathdesign}
\DeclareSymbolFont{usualmathcal}{OMS}{cmsy}{m}{n}
\DeclareSymbolFontAlphabet{\mathcal}{usualmathcal}

\begin{document}

\begin{center}{\Large \textbf{
The study of hadron properties and structure in high energy multi-production processes
}}\end{center}

\begin{center}
Shi-Yuan Li\textsuperscript{*},
\end{center}

\begin{center}
 Institute of Theoretical Physics, Shandong University
\\
* lishy@sdu.edu.cn
\end{center}

\begin{center}
\today
\end{center}

%

\section*{Abstract}
{\bf
In high energy scattering, the multi-production process is unique in its relevance to the total cross section and in its global properties, such as rapidity and other kinematic distributions. If there is a hard interaction, the jet rate and structure is a good arena for perturbative quantum chromodynamics. However, once any  hadron is specified, its properties and structure  must make sense, while the global and/or perturbative chromodynamic mechanism still can put important constraints. The relation of the  hadron's properties and structure  with its production cross section, distribution etc. can be  much more complex than its decay width. On the one hand there are many difficulties and challenges in the calculations, while  on the other hand, the production process provides a unique way to study the details of properties and structure of the hadron, which is beyond the approach of its decay process. Here I  review our works on such topics in recent years, mainly on the multi-quark state  production in multi-production processes and the Bethe-Salpeter wave function in exclusive processes. For the former, I  emphasize the unitarity of the hadronization process and relevant models,  so that almost no multi-quark state is  observed  in multi-production processes. I  also address how to calculate hadron molecule production in multi-production processes. The most recently observed  $T^+_{cc}$ is also stated with its relevance to colour and baryon number fluctuation of the preconfinement clusters. For the latter, I emphasize the Dirac structure of the hadron-quark coupling vertex.
}

\vspace{10pt}
\noindent\rule{\textwidth}{1pt}
\tableofcontents\thispagestyle{fancy}
\noindent\rule{\textwidth}{1pt}
\vspace{10pt}

\section{Introduction}
\label{sec:intro}
This symposium is about multiple particle dynamics.  Multiple particle systems are mainly produced from the high energy scatterings, of which the multi-production process is a dominant part of the total cross section. It is also the reason for the increasing total cross section of hadronic scattering \cite{cw}.  This topic, as well as other global properties, jet observable, etc. are all important subjects in studying QCD in high energy scattering. At the same time,  multi-production process is also a copious source of various hadrons. This last fact indicates that  multi-production process is a good arena to study hadron structure.
Compared to other ways of studying hadron structure, such as   some static property if available,  decay width  distributions, 'projector/probe' scattering  (e.g. DIS) etc., this way is more complex. However this is why this study  provides more complex information, even with different physical pictures. On the  one hand, there are many difficulties and challenges in the calculations, while on the other hand, the production process provides a unique way to study the details of the properties and structure of the hadron, which is beyond the approach of its decay process. This study relates the production mechanism with the structure and properties of a specific hadron.
Our study has two aspects: one is  the global properties, putting constraints on the production mechanism of  hadrons;  the other  is looking at the concrete details of the structure of   certain hadrons.
%
%
\section{Unitarity,  colour and baryon fluctuation}
Prompt  production of the  multi-quark states and/or  bound states of other ingredient hadrons  in high energy scattering can set a crucial bench-mark for  understanding the hadronization mechanism, since they contain more than three constituent quarks. In any hadronization process, the produced color-singlet (anti)quark system (e.g., preconfinement cluster  \cite{AV}) eventually transitions to various hadron states (mesons, baryons and beyond) with a total probability of exactly 1, which reflects the fact that there are no free quarks in the final states of any high energy process (confinement).   This process conserves entropy, since it only makes a  unitary transition on the density matrix. This consideration is very important, since such an analysis releases and shuts down a long paradox that  the entropy will decrease  and  introduce unphysical predictions in a combination process/model. We have also pointed out that  energy conservation and  unitarity of the  combination model are enough for this physical requirement.
The introduction of multi-quark states sets a challenge for the hadronization models/the mechanism dealing with the transition from color-singlet (anti)quark system to the hadron system. With these new  states introduced, a more detailed investigation of the whole hadron Hilbert space as well as that of the quarks is needed.
As a matter of fact, from  experiments, the production of general mesons and baryons is dominant, so the production rate of exotic particles  could be  small if  nonvanishing. However,
the present knowledge is not enough to judge how many kinds of multi-quark states there are and how they `share' the total probability, $\epsilon$. It is  not easy to predict the production rate of a specific multi-quark state. What we can say, though, it  that if there are many kinds of multi-quark hadrons, each only shares a small part of the small $\epsilon$. So the production rate of each is almost vanishing. This is consistent with the fact almost no multi-quark state  observed in multi-production processes.
The exotic hadrons  are observed to be produced from the decays of heavier hadrons (e.g. bottom hadrons) rather than promptly produced from multi-production processes in experiments.  From the theoretical aspects, this fact is understood, not only  because of the unitarity constraint  but also the modest mass of  the preconfinement clusters \cite{Han:2009jw,Jin:2016vjn,Jin:2010yd,Li:2019vrc}, which is independent from the collision  energy, and  results from the interplay between perturbative and non-perturbative QCD.
So to understand this topic, the preconfinement concept \cite{AV} is also very important, especially
in the case of a large number of quarks produced (e.g. in high energy nuclear collisions). This is consistent with unitarity, confinement, etc.  Furthermore one has to consider the fluctuations of colour and baryon number besides the hadronization models.  These fluctuations can be that of other quantum number, e.g. stangeness.
The relation lies in that all kinds of  multi-quark state hadrons have one common property,  that
the bound (anti)quarks inside can be grouped into several clusters, with each
cluster {\it possibly} in colour-singlet.  But the ways of grouping
these (anti)quarks are not unique;
 and dynamically, the colour interactions in
the system via exchanging gluons can change the colour state of each
individual  cluster, so each method of grouping/reduction  seems to have no special physical
reason. This  ambiguity has been discussed in other circumstances, under the  name of   'colour
recombination/rearrangement' \cite{our1, our2, Jin:2013bra}.   This fact shows that
multi-quark hadron can not be considered  in a unique and uniform way.
This is a  phenomenological
duality:  even the production of multi-quark hadrons is considered as
'hadron molecule formation' ('production definition'), however, the subsequent colour interactions
 in the system can eventually transit  this
'molecule' into a 'real'  multi-quark hadron, at least by some
probability --- and vice verse \cite{Jin:2014nva, Jin:2015mla,Li:2020ggh, lsy2016,r2005lsy}. The baryon number fluctuation means some cluster can have  one or more extra $qqq (\overline{qqq})$.  Based on this consideration, the colour and baryon fluctuation of the preconfinement clusters has non-trivial relevance to the multi-quark hadrons, and can be applied to construct various models for a specific multi-quark state for comparison.
\section{Bethe-Salpeter  wave function in production}
 An example is to investigate the exclusive production ratio $\frac{\sigma(e^+e^-\to K_S K_L)}{\sigma(e^+e^-\to K^+K^-)} $ in the  $e^+e^-$  continuum below the mass of $J/\Psi$,  under the spirit of Straton Model (i.e. completely relativistic Bethe-Salpeter  framework).   The coupling of the virtual photon to the Kaons is via a triangle quark loop: the photon-quark-quark vertex is exactly that of   Standard Model. The vertices between the quarks and the corresponding Kaon is the  Bethe-Salpeter vertex, in terms of the valence quark field. Hence the electromagnetic interaction and non-perturbative QCD interaction are  separately assigned. The difference of these two kinds of channels  lies in the electric charge.  Due to  the  scalar wave function    in the Bethe-Salpeter  vertex, which can be considered to regularize and renormalize the infinite integrations, the loop integral is finite.  The ratio  can be calculated straightforwardly \cite{Jin:2019gfa}, by adopting the vertex as $\gamma^5 (1+ B_1 \gamma_\mu P^\mu/M) \phi( q^2)$ \cite{Bhatnagar:2005vw}.  One gets   
$\frac{\sigma(e^+e^-\to K_S K_L)}{\sigma(e^+e^-\to K^+K^-)} \cong (\frac{m_s-m_d}{M})^2$. 
This is consistent with former experiments and can be further tested by BESIII measurements. The method was also applied to exclusive production process $e^+e^- \to J/\Psi+ \eta_c$ and explains the  data well.
\section{Conclusion}
To study the structure of hadrons via production is an important.  The recently observed $T_{cc}$ \cite{LHCb:2021vvq,LHCb:2021auc} is investigated taking into account the colour connection and baryon number fluctuation \cite{Li:2019vrc}, and a compact four-quark nature is favoured \cite{Jin:2021cxj}. The interplay of multi-HEAVY-quark production and multi-HEAVY-quark bound states is a quite new and copious direction in this field. Now three pairs of charm can be carefully investigated in LHC (see, e.g., \cite{CMS:2021qsn}).
\section*{Acknowledgements}
I thank all Collaborators.
This work is supported in part by  National Natural Science Foundation of China (grant Nos.  11635009, 11775130).


\begin{thebibliography}{150}
\bibitem{cw}
H. Cheng and T. T. Wu, Expanding Protons: Scattering at High Energies, MIT Press, Cambridge, MA, 1987.
 \bibitem{AV}
 D. Amati and G. Veneziano, Phys. Lett. B 83 (1979), 87-92.

\bibitem{Han:2009jw}
W.~Han, S.~Y.~Li, Y.~H.~Shang, F.~L.~Shao and T.~Yao,
Phys.\ Rev.\ C, {\bf 80}, 035202 (2009).
 \bibitem{Jin:2016vjn}
Y.~Jin, S.~Y.~Li, Y.~R.~Liu, L.~Meng, Z.~G.~Si and X.~F.~Zhang,
Chin. Phys. C \textbf{41} (2017) no.8, 083106
doi:10.1088/1674-1137/41/8/083106
[arXiv:1610.04411 [hep-ph]].
\bibitem{Jin:2010yd}
Y.~Jin, S.~Y.~Li, Z.~G.~Si and T.~Yao,
[arXiv:1005.4664 [hep-ph]].
\bibitem{Li:2019vrc}
S.~Y.~Li,  Moriond QCD 2019, 107-110.


  \bibitem{our1}
 W.~Han, S.~Y.~Li, Z.~G.~Si and Z.~J.~Yang, Phys. Lett. B {\bf 642}, 62-67 (2006).
 \bibitem{our2}
 Z.~G.~Si, Q.~Wang and Q.~B.~Xie,
  Phys.\ Lett.\  B {\bf 401} (1997) 107;
  Q.~Wang, Q.~B.~Xie and Z.~G.~Si,
  Phys.\ Lett.\  B {\bf 388} (1996) 346.

\bibitem{Jin:2013bra}
Y.~Jin, S.~Y.~Li, Z.~G.~Si, Z.~J.~Yang and T.~Yao,
Phys. Lett. B \textbf{727} (2013), 468-473
doi:10.1016/j.physletb.2013.10.070
[arXiv:1309.5849 [hep-ph]], PHYSICAL REVIEW D 89, 094006 (2014).

\bibitem{Jin:2014nva}
Y.~Jin, S.~Y.~Li, Y.~R.~Liu, Z.~G.~Si and T.~Yao,
Phys. Rev. D \textbf{89} (2014) no.9, 094006
doi:10.1103/PhysRevD.89.094006
[arXiv:1401.6652 [hep-ph]].

\bibitem{Jin:2015mla}
Y.~Jin, H.~L.~Li, S.~Q.~Li, S.~Y.~Li, Z.~G.~Si, T.~Yao and X.~F.~Zhang,
Phys. Rev. D \textbf{91} (2015) no.11, 114017
doi:10.1103/PhysRevD.91.114017
[arXiv:1506.00121 [hep-ph]].

\bibitem{Li:2020ggh}
S.~Y.~Li, Z.~Y.~Li, Z.~G.~Si, Z.~J.~Yang and X.~Zhang,
[arXiv:2007.07706 [hep-ph]].





 \bibitem{lsy2016}
Y.~Jin, S.~Y.~Li and S.~Q.~Li,
Phys. Rev. D \textbf{94} (2016) no.1, 014023
doi:10.1103/PhysRevD.94.014023
[arXiv:1603.03250 [hep-ph]].

 \bibitem{r2005lsy}
  S. Y. Li, Calculation of production and decay of baryonium via positronium (quarkonium) approach, ICTP Visitor's Work Report 2005 (unpublished).

\bibitem{Jin:2019gfa}
Y.~Jin, S.~Y.~Li, Y.~R.~Liu, Z.~X.~Meng, Z.~G.~Si and T.~Yao,
Phys. Rev. C \textbf{102} (2020) no.1, 015201
doi:10.1103/PhysRevC.102.015201
[arXiv:1906.11575 [hep-ph]].

\bibitem{Bhatnagar:2005vw}
S.~Bhatnagar and S.~Y.~Li,
J. Phys. \textbf{32} (2006), 949-961
doi:10.1088/0954-3899/32/7/005
[arXiv:hep-ph/0512352 [hep-ph]].

\bibitem{LHCb:2021vvq}
R.~Aaij \textit{et al.} [LHCb],
[arXiv:2109.01038 [hep-ex]].

\bibitem{LHCb:2021auc}
R.~Aaij \textit{et al.} [LHCb],
[arXiv:2109.01056 [hep-ex]].

\bibitem{Jin:2021cxj}
Y.~Jin, S.~Y.~Li, Y.~R.~Liu, Q.~Qin, Z.~G.~Si and F.~S.~Yu,
[arXiv:2109.05678 [hep-ph]].

\bibitem{CMS:2021qsn}
A.~Tumasyan \textit{et al.} [CMS],
[arXiv:2111.05370 [hep-ex]].




\end{thebibliography}
\end{document}